\documentclass[times,twocolumn,final,authoryear]{elsarticle}

\usepackage{jasr}
\usepackage{framed,multirow}

\usepackage{amssymb}
\usepackage{latexsym}

\usepackage[switch]{lineno}

\usepackage{url}
\usepackage{xcolor}
\definecolor{newcolor}{rgb}{.8,.349,.1}

\usepackage[citebordercolor=white]{hyperref}

\journal{Advances in Space Research}

\begin{document}

\verso{D. Kang \textit{et al.}}

\begin{frontmatter}

\title{Recent results of cosmic-ray studies with IceTop at the IceCube Neutrino Observatory}

\author[1]{Donghwa \snm{\textcolor{black}{Kang}}\corref{cor1}}
\cortext[cor1]{Corresponding author:
  donghwa.kang@kit.edu
}
\author[]{on behalf of the IceCube Collaboration}

\address[1]{Karlsruhe Institute of Technology, Institute for Astroparticle Physics, Karlsruhe, Germany}

\received{6 April 2023}
\finalform{4 September 2023}
\accepted{12 September 2023}
\availableonline{}
\communicated{}

\begin{abstract}
The IceCube Neutrino Observatory is a cubic-kilometer Cherenkov detector that is deployed deep in the Antarctic ice at the South Pole. A square kilometer companion surface detector, IceTop, located directly above in the in-ice array, measures cosmic-ray initiated extensive air showers with primary energies between 100~TeV and 1~EeV.
By combining the events measured by IceTop and the in-ice detectors of IceCube in coincidence, we can reconstruct the energy spectra for different primary mass groups. Therefore, we provide information about the origin of cosmic rays, in particular, in the transition region from galactic to extra-galactic origin of high-energy cosmic rays. In this contribution we present recent experimental results, as well as prospects by the foreseen enhancement of the surface detectors of IceTop and the future IceCube-Gen2 surface array.
\end{abstract}


\end{frontmatter}


\section{Introduction}
\label{sec1}
Investigations of the energy spectrum and mass composition of primary cosmic rays (CR) are fundamental tools to understand the origin, acceleration and propagation mechanism of these particles. It is also important for the determination of the atmospheric neutrino flux. The presumed origin of galactic cosmic rays are supernovae. The shock acceleration at supernova remnants can explain the intensity of the cosmic radiation at least up to 10$^{15}$~eV \citep{Hillas2005}.
A continuous and steady source distribution would generate an energy spectrum of cosmic rays with a simple power law for all the elements. However, in a more realistic approach, sources are discrete and a possible non-uniform distribution in space and time could generate structures and changes in the spectral indexes of the spectra of primaries at certain energies \citep{Peters1961}. This situation would be more pronounced at higher energies, where the most recent sources would dominate the spectra.
A refined study of the CR primary spectrum, composition and anisotropy is, therefore, extremely important to address the open questions: the astrophysical origin of cosmic rays, and conditions at the acceleration sites.

The direct study of CRs can be performed by means of satellite or balloon-borne detectors only at energies below 10$^{15}$~eV \citep{CREAM2011}.
Above that energy, due to statistics, only Extensive Air Showers (EAS) detection can be used \citep{KASCADE2005}.
By using multiple particle detectors to study different air shower components, the energy of the primary particles can be inferred in a quite reliable way through the comparison of the data with extensive simulation studies. However, the results of the composition analysis depend to some extent
on the high-energy interaction model used in the simulations.

Above 10$^{15}$~eV the all-particle spectrum has a power-law-like behavior ($dN/dE \propto E^{\gamma}$, $\gamma \approx -2.7$) with features known as the knee around 3-5$\times$10$^{15}$~eV and the ankle at 4-10$\times$10$^{18}$~eV, respectively, where the spectrum shows a distinct change of the spectral index \citep{Bluemer2009}.
The energy range between 10$^{17}$~eV and 10$^{19}$~eV is in particular very interesting as it is expected to be the region where a transition from a galactic dominated to an extra-galactic dominated origin is observed \citep{Hillas2005}.
The Pierre Auger collaboration confirmed the extragalactic origin of cosmic rays beyond the ankle \citep{AugerScience2017}, but not the transition itself. The exact energy transition region from galactic to extragalactic cosmic rays is not well understood yet.
Therefore, the study of the chemical composition and of the shape of the energy spectrum in this energy range is also of great interest.

The muon content of EAS is important for identifying properties of primary cosmic rays. Since the muonic component in an air shower depends on the energy and the mass of the primary particle, it plays an important role in studies of the cosmic-ray composition.
They carry information about the last hadronic interaction that created their parent pion.
Due to the propagation of muons at nearly the speed of light, this information is reflected in the lateral density profile and in the arrival times of muons when an air shower reaches the ground.
Several experiments have observed the muon lateral density profile and found a consistent deficit in the simulated numbers of muons in the lateral profile compared to experimental data. This discrepancy is referred to as the muon puzzle \citep{Albrecht2022}.
IceCube (Fig.~1) has a unique capability to measure the GeV and TeV muon components in air showers, separately.
Therefore, measurements by IceTop and IceCube can give a clue on the muon puzzle.

Recent findings indicate that one of the most promising methods to bring further insights into high-energy CR physics is the observation of CR-initiated EAS using multiple detection channels. The IceCube Neutrino Observatory constitutes a unique setup, allowing for in-ice detection of the high-energy core of the air-shower as well as its footprint using IceTop.

\section{The IceCube detector}
The IceCube Neutrino Observatory \citep{IceCube2017} is a cubic-kilometer Cherenkov detector that is deployed deep in the Antarctic ice at the South Pole.
The schematic view of the IceCube Neutrino Observatory can be seen in Fig.~1.
It has 5160 digital optical modules (DOMs) placed on 86 strings in the glacial ice at depths between roughly 1450~m to 2450~m below the surface.
Each DOM consists of a 25~cm photomultiplier tube plus associated digitization and calibration electronics, hosted inside a 35~cm diameter pressure vessel.
Reconstruction of the direction, energy and identity of penetrating particles at IceCube relies on the optical detection of Cherenkov radiation emitted in the surrounding ice.
In addition to the in-ice array, a square kilometer companion surface detector, IceTop with a mean spacing of 125~m \citep{IceTop2013}, located directly above the in-ice array, measures extensive cosmic ray air showers with primary energies between 100~TeV and 1~EeV.
The altitude of the surface array is 2835 m a.s.l., which corresponds to an atmospheric depth of about 680~g/cm$^{2}$.
It is close to the shower maximum for showers with tens of PeV energies, therefore, an excellent energy resolution is expected at these primary energies. IceTop consists of 81 stations with two Cherenkov tanks each and each tank is equipped with two Digital Optical Modules (DOMs) identical to those used by IceCube in-ice array.
During deployment, water was filled into the tanks and allowed to freeze under controlled conditions to ensure bubble-free, clear ice.
The angular resolution of air showers detected by IceTop is about 1$^{\circ}$ and the timing resolution is around 3~ns.
The energy resolution is about 0.1 or less in units of log$_{10}$(E/GeV) above about 1 PeV \citep{PRD2019}.

\begin{figure}
  \centering
  \includegraphics[scale=0.3]{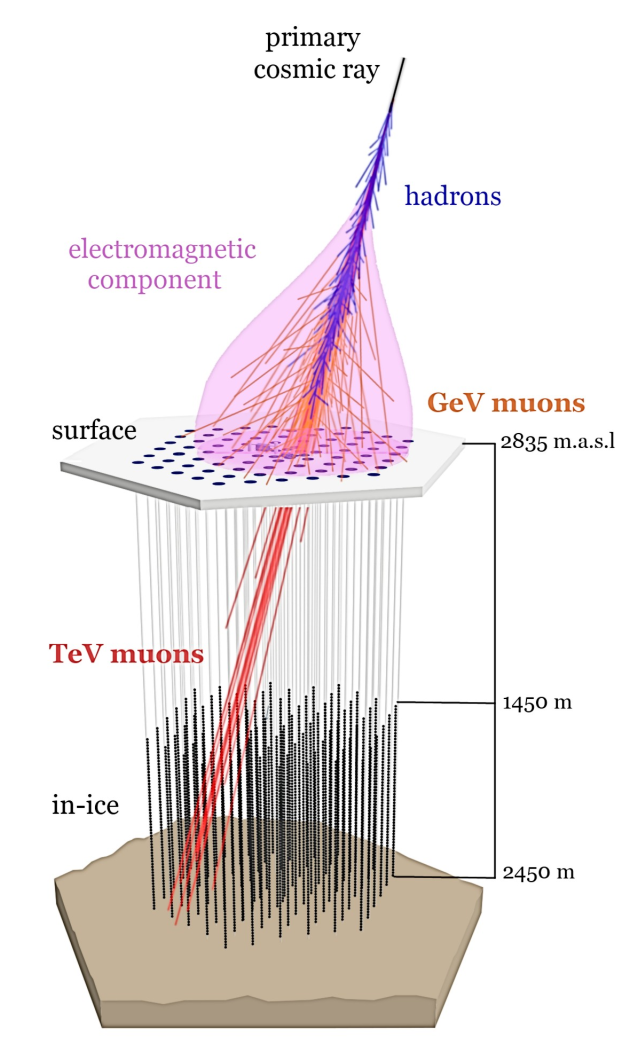}
  \caption{Schematic view of an air shower observed with the IceCube Neutrino Observatory \citep{Verpoest2021}.}
\end{figure}

IceCube is a unique instrument for cosmic ray physics, by using a three-dimensional detector concept (Fig.~1).
The surface array IceTop measures the electromagnetic and low energy ($E_{\mu} \sim$ 1~GeV) muon components of extensive air showers.
From that, the energy and the direction of cosmic rays are reconstructed.
The high-energy muons ($E_{\mu} >$ 400~GeV) go through and can be measured by the in-ice detector.
The track or bundle of muons is reconstructed and the deposited energy along the track is used as an in-ice energy proxy for the analysis.
Using both information of IceTop and in-ice, composition studies can be performed more precisely.

\section{Energy spectrum}
\subsection{IceTop alone}
In the ``IceTop-alone'' analysis, the energy spectrum is reconstructed using only data from the IceTop tanks. Although it does not use accompanying information from the in-ice detector, it can access a greater range of zenith angles and larger statistics.
The tank signals collected by IceTop are cleaned and calibrated in units of vertical equivalent muons (VEM), and these cleaned data are reconstructed \citep{IceTop2013}.
The unit 1~VEM is defined as the charge value at 95\% of the muon peak value, which is the maximum of the muon contribution \citep{IceTop2013}.
The properties of the primary cosmic ray are reconstructed by fitting the measured signals with a Lateral Distribution Function (LDF), which depends on the perpendicular distance from the shower axis, $r$:
\begin{eqnarray}
S(r) = S_{125} \cdot (\frac{r}{125{\rm m}})^{-\beta-\kappa \cdot {\rm log}_{10}(r/125{\rm m})},
\end{eqnarray}
where $S_{125}$ is the shower size at 125~m from the shower axis and $\beta$ is a slope of the LDF function and a free parameter.
The curvature of the parabola, $\kappa$, is fixed at a value of 0.303, which studies suggest is similar across different hadronic interaction models.

The fitting of the lateral distribution of charge signals results in the shower size parameter $S_{125}$, expressed in the unit of VEM, which is used as an energy proxy of IceTop. This parameter is nearly composition independent.

Figure\ 2 presents the correlation between $S_{125}$ and the primary energy in IceTop for the primary proton assumption for cos$(\theta) >$ 0.95.
Events are weighted to the relative abundances of the primary nuclei in the mixed composition H4a model \citep{Gaisser2012}.
For the Monte Carlo simulation, the hadronic interaction model Sibyll~2.1 \citep{Ahn2009} is used, whereas FLUKA \citep{FLUKA2005} is used for low energy hadronic interactions below 80 GeV.

\begin{figure}
  \centering
  \includegraphics[scale=0.21]{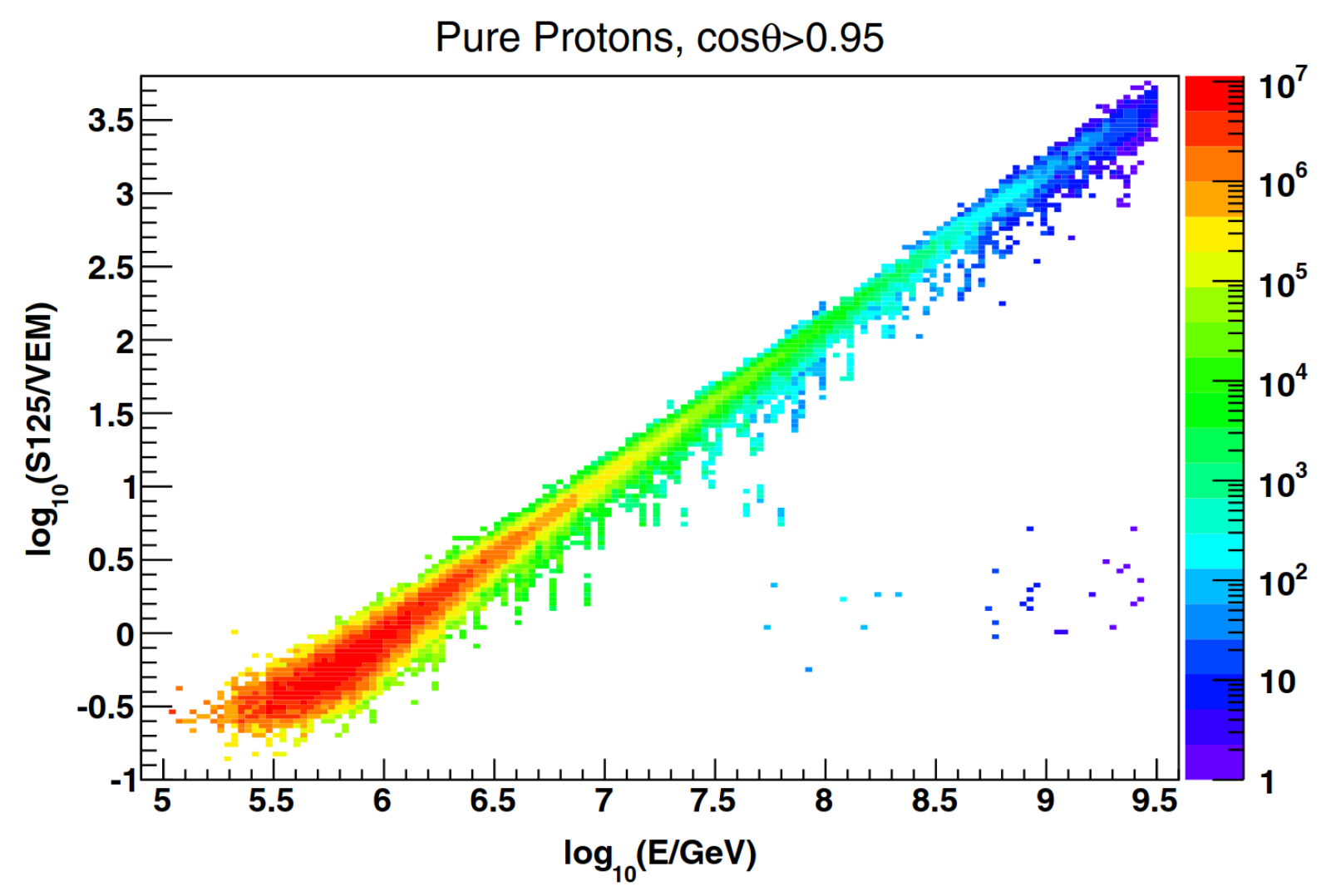}
  \caption{Shower size $S_{125}$ as a function of the true primary energy in simulated air showers assuming a proton primary for small zenith angles \citep{PRD2019}. Although quality cuts eliminate most of the misreconstructed events in the data sample, the blob in the lower right represents the small number of them that do survive in the sample.
  }
\end{figure}

By means of the relationship between $S_{125}$ and the primary energy, a conversion function \citep{PRD2019} was developed, which depends on the composition assumption, the zenith angle and the assumed spectral index. After applying this relationship to experimental data, the all-particle energy spectrum is reconstructed.

For this analysis three years of data from May 2010 to June 2013 are used, and the selected events after considering quality cuts and full
trigger and reconstruction efficiency is $5 \cdot 10^{7}$ in total.
Most of the IceTop selections rely on the success of the reconstruction algorithm, while the in-ice quality selections focus on ensuring an accurate reconstruction of the energy loss. Detailed cuts for the analysis are described in Ref.~\citep{PRD2019}.
The resulting spectrum derived from three years of data is displayed in Fig.~3, where the bin-to-bin migration effect is expected, but it is so small as to be negligible.
It shows two features, the so-called knee around 5~PeV and a second knee around 100~PeV.
The observation of the second knee structure is relatively recent, but it is now confirmed by at least three different experiments: KASCADE-Grande, IceCube, and TUNKA \citep{PRL2011, PRD2019, TUNKA2020, Auger2021}.

\begin{figure}
  \centering
  \includegraphics[scale=0.21]{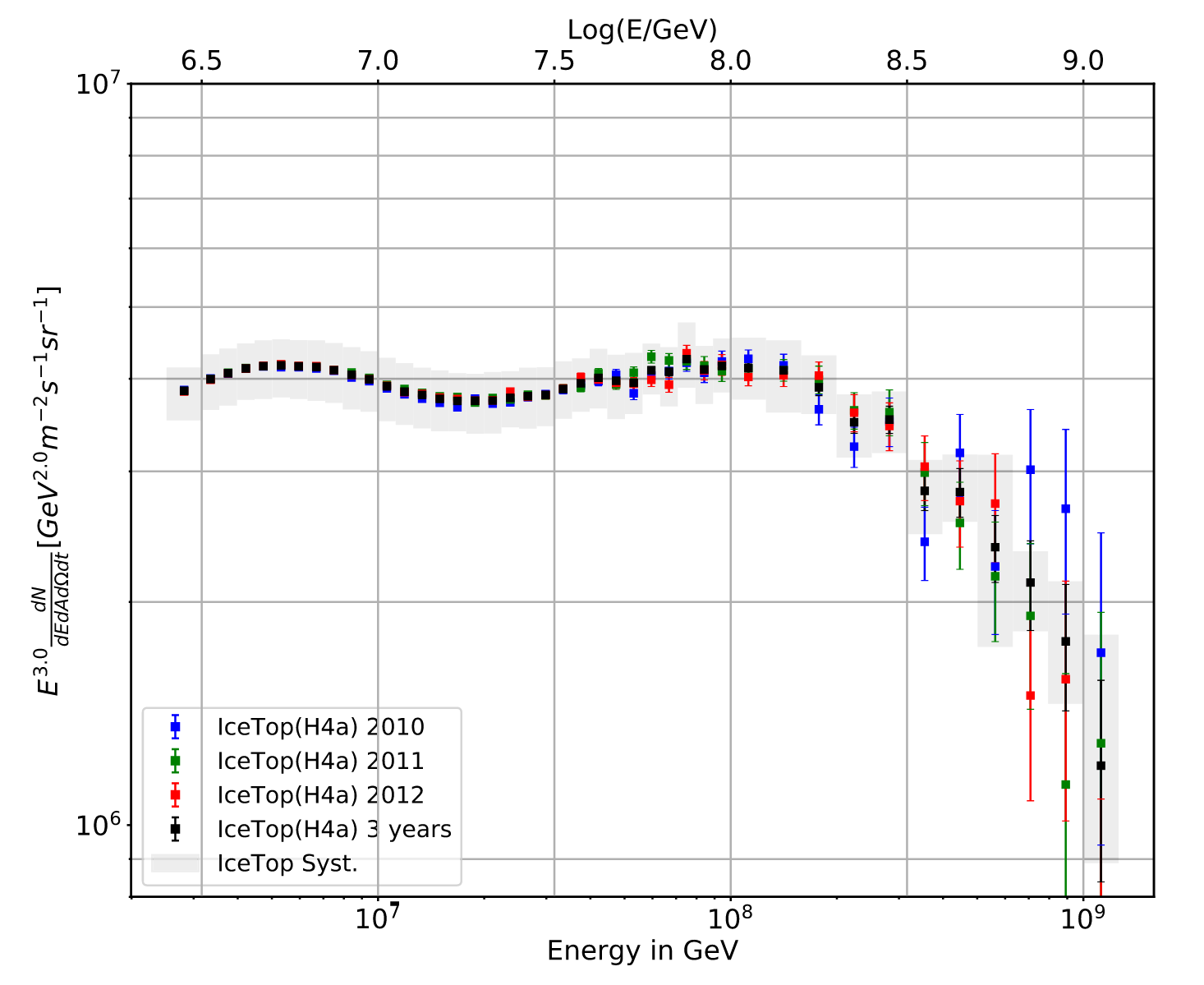}
  \caption{All-particle flux of cosmic rays obtained from three years of IceTop data \citep{PRD2019}, derived using the H4a mass composition assumption. Spectra for each individual year are shown as well. The gray band indicates the systematic uncertainty of the flux. The error bars are statistical uncertainties.}
\end{figure}

An important systematic effect is the varying snow layers on top of the IceTop tanks. There is considerable snow drift at the South Pole and it leads to snow accumulation of an average of 20\ cm per year on top of the IceTop array. Due to snow effects, the electromagnetic components are attenuated and this effect changes slowly over time. Thus, the attenuation by snow is corrected for using a simple exponential factor \citep{Rawlins2013}:
\begin{eqnarray}
  S_{corr} = S_{0} \cdot \exp(-X/\lambda_{s}),
\end{eqnarray}
where $S_{0}$ is the no-snow expected signal, $X = d_{snow}/{\rm cos}(\theta)$ is the slant depth that particles travel to a tank at a depth of $d_{snow}$, $\lambda_{s}$ is
a year-dependent effective attenuation length between 2.1 and 2.3 meters.

The dataset divided into individual years shows strong agreement with each other, with systematic uncertainties of about 10\%, where the dominant contributions are from snow accumulation. Detailed discussions on the systematic uncertainties can be found in Ref.~\citep{PRD2019}.

\begin{figure*}
  \centering
  \includegraphics[scale=0.26]{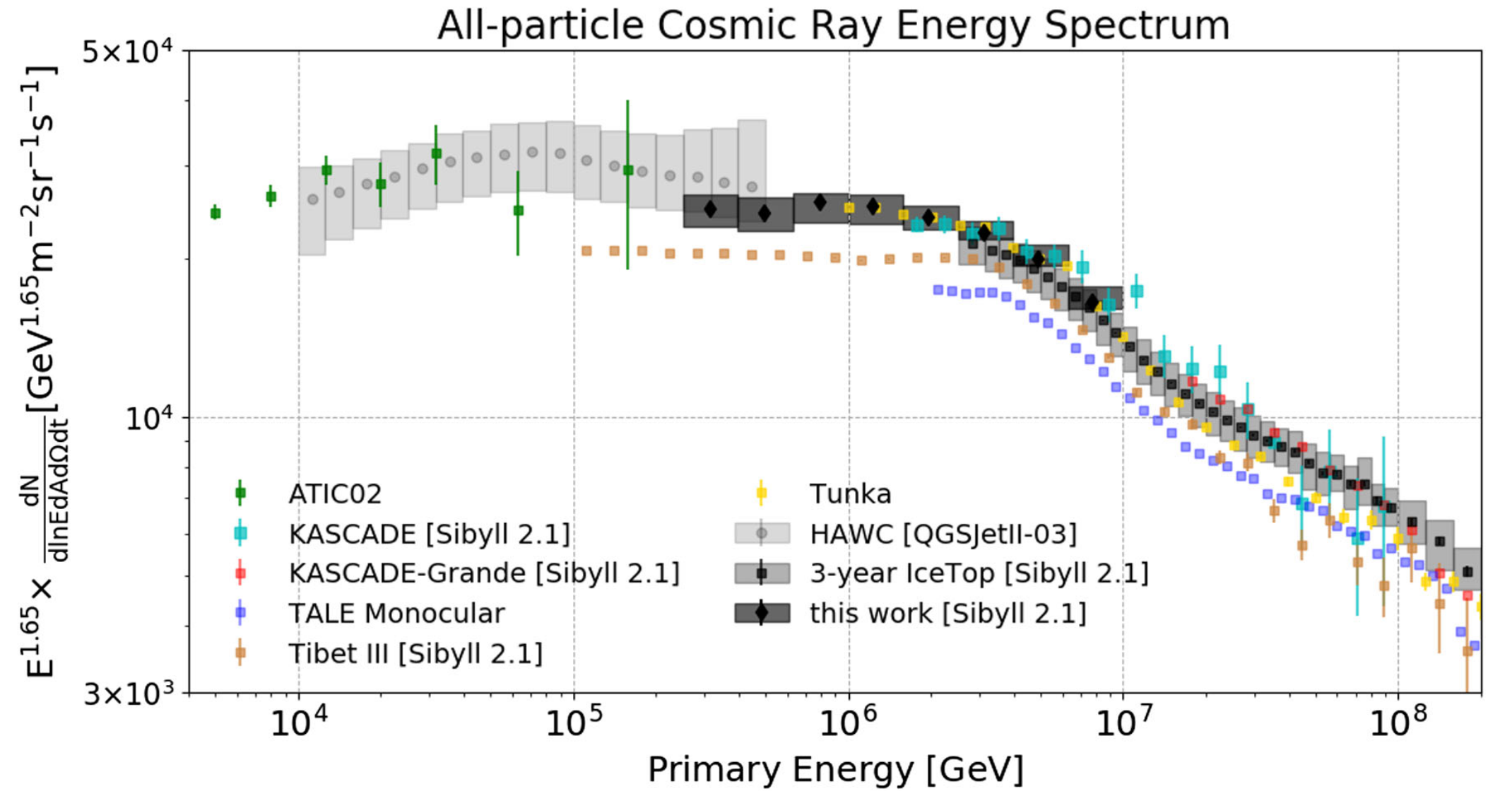}
  \caption{All-particle energy spectrum of cosmic rays using IceTop 2016 data scaled by $E^{1.65}$ \citep{PRD2020}. The energy spectrum from this analysis (black dots with dark gray band) is compared with the cosmic-ray flux inferred by other experiments including HAWC \citep{HAWC2017}, KASCADE \citep{KASCADE2005}, KASCADE-Grande \citep{PRL2011}, TALE \citep{TALE2018}, Tibet-III \citep{TibetIII2008} and Tunka \citep{TUNKA2020}.} 
\end{figure*}

\subsection{Low energy spectrum}
One of the important goals for the reconstruction of the primary energy spectrum of cosmic rays is to investigate the overlap region between direct and indirect measurements. Recently, IceCube has published a new low-energy spectrum \citep{PRD2020}, which is extended to low energies down to 250~TeV, using the IceTop infill array.
IceTop has several closely-spaced stations in the center of the array, for which the separation between stations is less than 50~m;
this densely-instrumented region is sensitive to cosmic rays with low energy.
A two-station trigger is implemented to collect lower energy events. This trigger condition is fulfilled when four pairs of infill stations are hit within a time window of 200~ns.
In this analysis, a random forest algorithm for reconstruction of showers was used, since there are insufficient hits for the standard fit to a lateral distribution function.
Each simulated event is weighted based on the H4a cosmic ray primary composition model during training to remove an energy-dependent bias on reconstructed energy for cos$(\theta) \leq$ 0.9.
The reconstructed energy distribution which was derived from the random forest regression is unfolded by using an iterative Bayesian unfolding procedure \citep{DAgostini1995} to take energy bin migration into account.

The total energy spectrum using IceTop 2016 data, scaled by $E^{1.65}$ and compared to the cosmic-ray flux from other previous experiments, is presented in Fig.~4. The spectrum shows a clear behavior in the knee region: a slope of 1.65 below PeV and a steepening between 2~PeV and 10~PeV.

The IceTop low-energy spectrum connects to direct measurements and overlaps with HAWC measurements \citep{HAWC2017} at lower energies,
where this and the analysis of the HAWC energy spectrum use the different hadronic interaction models Sibyll~2.1 and QGSJetII-03 \citep{Ostapchenko2006}, respectively.
The uncertainty of the HAWC spectrum appears larger than the one for IceTop. The reason is that the uncertainty from hadronic interaction models is included for HAWC, but not for IceTop.
In addition, the largest uncertainty for HAWC comes from uncertainties in the photomultiplier (PMT) parameters.
This result is also compared with the results of KASCADE and Tunka at higher energies. The Tibet-III measurement \citep{TibetIII2008} is the most relevant for comparison, since it is a ground-based air shower array at high altitude with small distances between detectors (7.5~m).
The Tibet-III result in Fig.~4 is based on the heavy-dominated (HD) composition model \citep{TibetIII2008} with Sibyll~2.1, whereas this analysis uses the composition model of H4a. Due to the different composition assumptions, an apparent difference (in energy scale of 20\%) between measurements appears.
The Telescope Array Low-Energy Extension (TALE) experiment detects low-energy cosmic rays in the PeV energy range using atmospheric fluorescence detectors, which are also sensitive to the Cherenkov radiation produced by shower particles.
There is a discrepancy between the TALE result \citep{TALE2018} and the ground based experiments due to systematic effects.
However, the low-energy spectrum shows an excellent agreement with the 3 years IceTop result in the overlapping region, where the knee structure is visible.

Due to the large statistics of data, the statistical uncertainty of the energy spectrum is small.
However, the total systematic uncertainties (dark shaded band) are rather large. The largest contribution to the uncertainty is the composition assumption, while the effects form the unfolding method, effective area and atmosphere are relatively small.
Total systematic uncertainties in percentage can be found in Ref.~\citep{PRD2020}. 
The effect of the hadronic interaction model is not included in the total systematic uncertainty.

\section{Mass composition}
\subsection{Coincident analysis}
Another way to estimate the energy spectrum was derived from coincident events seen in IceTop and IceCube.
For the coincident analysis, the behavior of the in-ice detector due to the high-energy muons is studied.
In-ice observables provide a handle on primary composition,
since the heavier the shower-initiating particle is, the more muons are expected for a given primary energy.
Therefore, iron-induced showers result in a larger deposit of Cherenkov light in IceCube than proton-induced showers of the same primary energy.
In addition, the energy loss of the muon bundle at a fixed slant depth depends strongly on the multiplicity of the muon bundle and accordingly on the composition.
Moreover, the stochastic behavior is also composition dependent since the probability of multiple muons emitting via a radiative energy loss on the same track is higher for iron, which has higher multiplicity, than for proton.
This technique was already exploited by the EAS-TOP and MACRO experiments where results on the mass composition in the knee region were derived \citep{ESATOP2004}.

Eventually, it turns out that proton showers are expected to create a small number of muons with higher-energy stochastic losses in the detector, while iron showers have more muons, but more lower-energy stochastic losses than proton showers. Thus, we consider this stochastic fluctuation as an additional composition-sensitive parameter. 

Figure\ 5 shows an average energy loss profile for a large event.
The energy loss profile is then fitted to obtain two composition-sensitive parameters: the average energy loss (red solid line in Fig.~5) and deviations from the average (i.e. stochastics).
Thus the energy loss parameter in ice, $dE_{\mu}/dX$, is defined as the value from the fitting to the energy loss profile at a fixed slant depth of 1500~m, which is roughly corresponding to the top of IceCube.
Two different selections of high-energy stochastics from an energy loss profile are used: a standard selection (red dashed line in Fig.~5) and a strong selection requiring higher stochastic energy losses (red dotted line in Fig.~5).

Figure\ 6 presents the correlation of energy proxies between the reconstructed shower size $S_{125}$ from IceTop and the reconstructed energy loss from IceCube for simulations with proton and iron assumptions. It shows a strong composition sensitivity, so there is a clear separation between proton and iron primaries.

\begin{figure}[h!]
  \centering
  \includegraphics[scale=0.25]{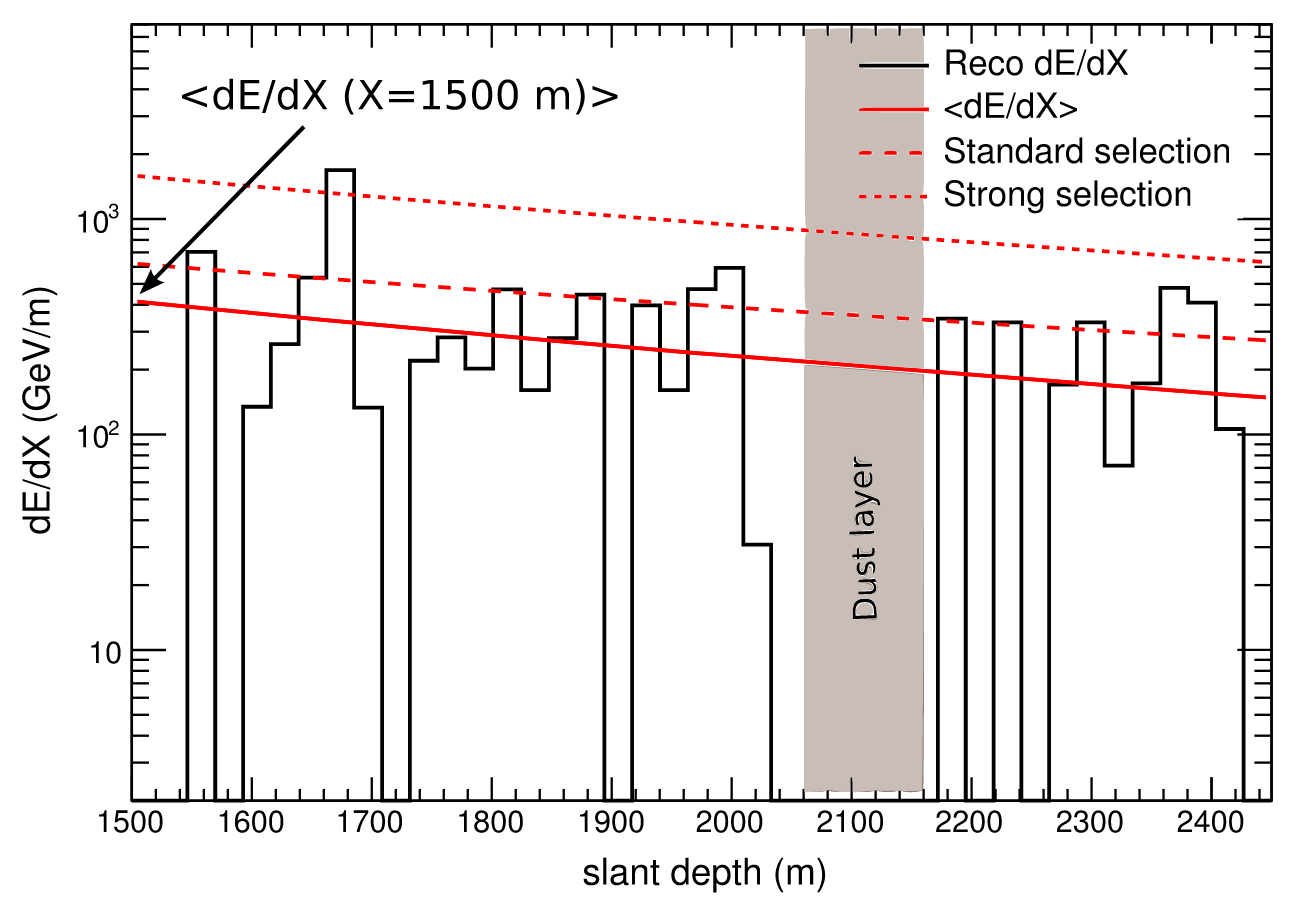}
  \caption{An example of a reconstructed in-ice energy loss profile from a large event \citep{PRD2019}. The solid red line indicates the average energy loss fit, the dashed red line presents the standard stochastics selection, and the dotted red line is the strong stochastics selection. The gray band shows the approximate location of the dust layer.}
\end{figure}

\begin{figure}
  \centering
  \includegraphics[scale=0.225]{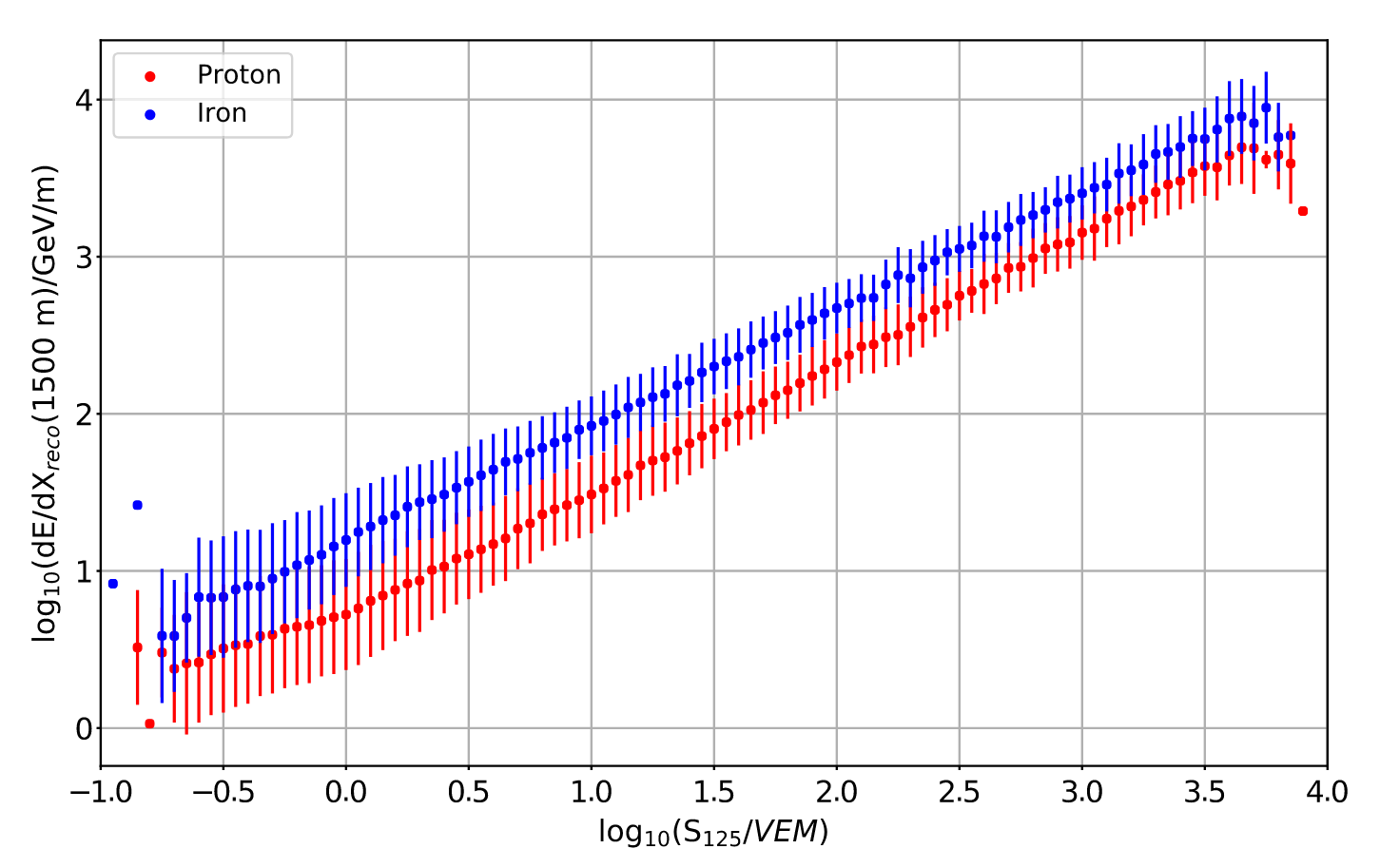}
  \caption{Reconstructed in-ice energy loss from IceCube as a function of $S_{125}$ for simulations of proton (red) and iron (blue) induced showers \citep{PRD2019}. Error bars indicate the standard deviation.}
\end{figure}

By means of coincident events of IceTop and IceCube, a neural network technique is applied to obtain the reconstructed primary energy and the primary mass.
This analysis uses five input parameters, which have a non-linear behavior in primary energy and mass.
The shower size $S_{125}$ and the zenith angle in terms of cos$\theta$ are from IceTop, whereas the energy loss of muons $dE/dX$ and the number of high-energy stochastics of standard and strong selections are from in-ice. The network has a strong dependency on the variables of $S_{125}$ and $dE/dX$. 
The neural network is trained on simulation to determine the correlation between the five inputs and the two outputs. Half of the sample of Monte Carlo simulation is used to train and test the neural network, while the other half is used for comparison of data and simulation to verify the sample.
The network was constructed from three groups of network structures and the number of neurons was varied within the hidden layers.
The neural network produces two outputs: one representing energy, and the other mass. The first energy output can be used to estimate the primary energy for each individual event, and to create a continuous energy distribution. The second mass output is a lower-resolution proxy for the primary mass number.
Therefore, for the mass composition, the mass is not classified on an event-by-event basis and further statistical analysis is required to decompose the primary mass.

\begin{figure}
  \centering
  \includegraphics[scale=0.27]{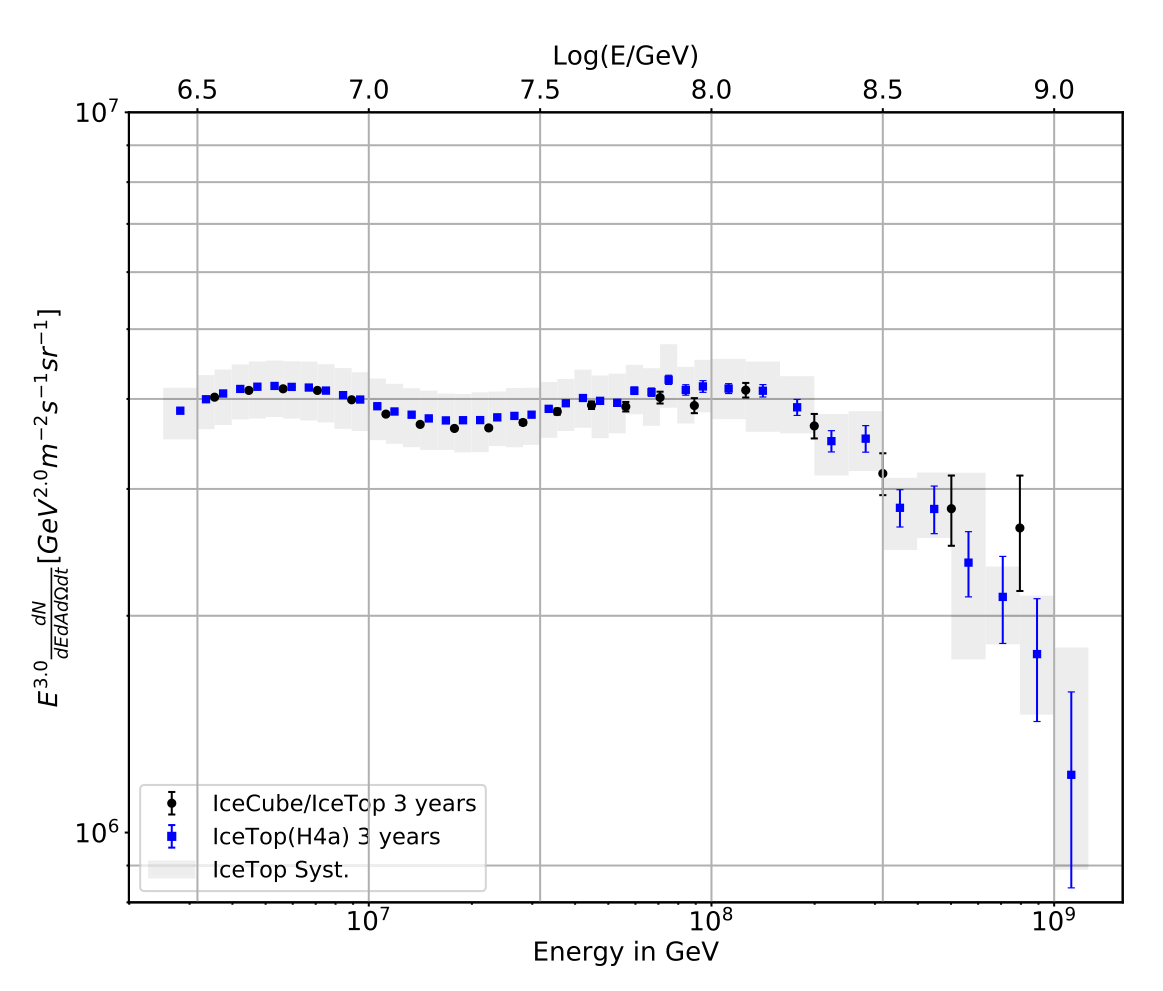}
  \caption{A comparison of the combined 3-year spectra \citep{PRD2019} based on the hadronic interaction model of Sibyll~2.1 from two different analyses: the IceTop-alone analysis (blue) and the coincident analysis (black). The gray band is the total systematic uncertainties of the IceTop detector. Error bars indicate statistical uncertainties.}
\end{figure}

Figure\ 7 displays a comparison of the energy spectra resulting from the IceTop-alone analysis (blue dots) and the coincident analysis (black dots) using the neural network technique. Both analyses are in good agreement with each other within statistical and systematic uncertainties. 

\begin{figure}
  \centering
  \includegraphics[scale=0.28]{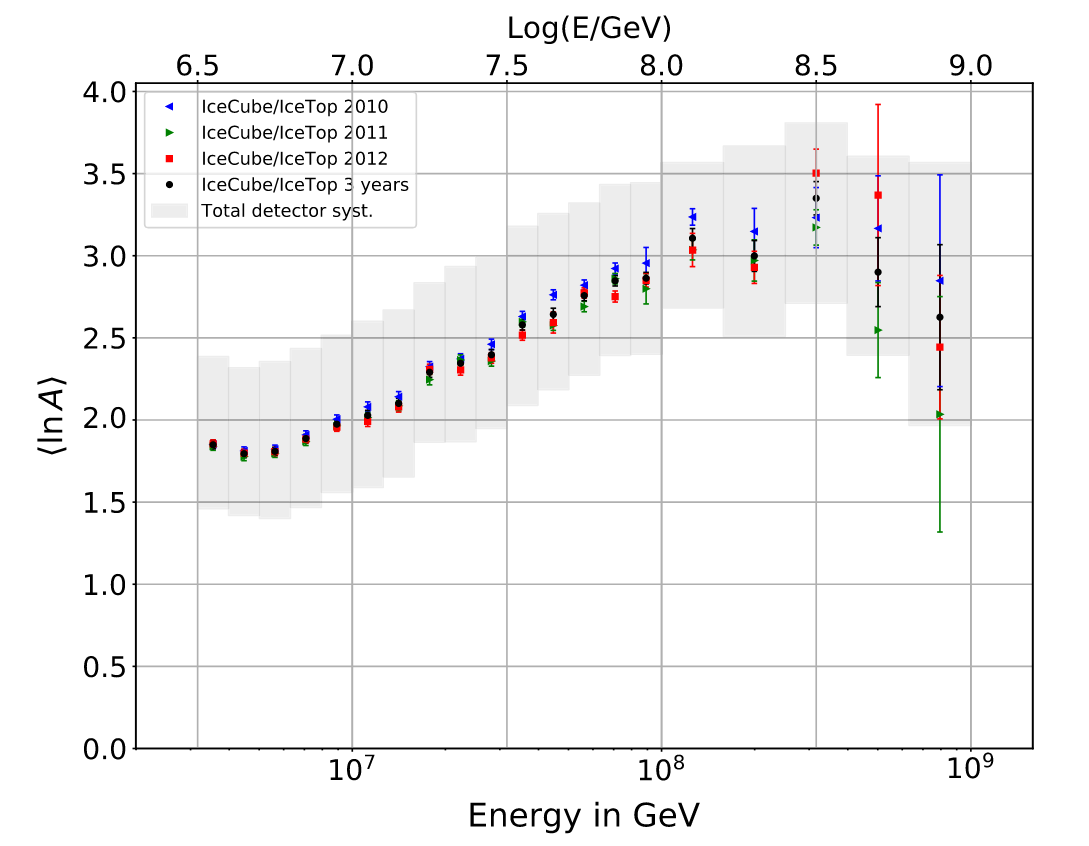}
  \caption{Mean logarithmic mass $\langle \ln A\rangle$ derived from the best fit to the neural network mass output for the 3 years of data and the individual years \citep{PRD2019}. The gray band represents the combined systematic uncertainties of the IceTop and in-ice detectors for the coincident analysis. Sibyll~2.1 was used for the hadronic interaction model.}
\end{figure}

The neutral network mass output in each energy bin for each primary element is converted into a template probability density function (p.d.f.) by using an adaptive kernel density estimation (KDE) method. The four template p.d.f.'s were weighted to find the fractions which best fit the neural network mass output for the experimental data in the same bin in reconstructed energy.

Figure\ 8 shows the mean logarithmic mass $\langle \ln A\rangle$, which is derived from the individual fractions. The gray band represents the total coincident detector uncertainties from both the IceTop and in-ice detectors, where the largest systematic contribution is from in-ice light yield of 10\%.
The distribution of the mean logarithmic mass shows a clear trend toward heavy nuclei with increasing energy.

\begin{figure*}[h]
  \centering
  \includegraphics[scale=0.33]{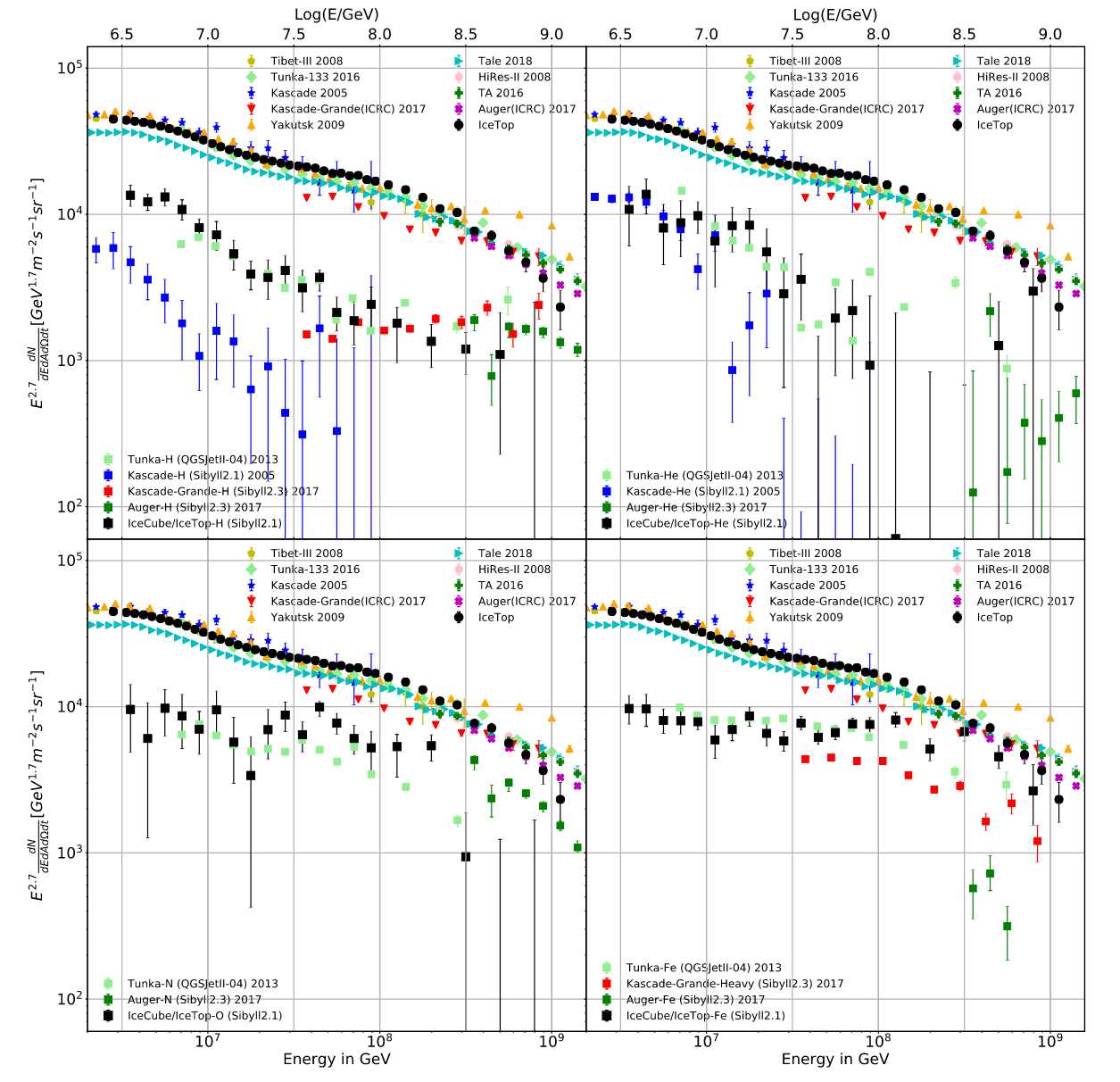}
  \caption{Comparison of the all-particle and individual energy spectra of the four mass groups - protons, helium, oxygen and iron - with other previous experiments \citep{PRD2019}. The hadronic interaction model of Sibyll~2.1 was used. The all-particle spectra of previous measurements are taken from Tibet \citep{TibetIII2008}, Tunka \citep{TUNKA2020}, KASCADE \citep{KASCADE2005}, KASCADE-Grande \citep{PRL2011}, Yakutsk \citep{Yakutsk2009}, TALE \citep{TALE2018} and Pierre Auger Observatory \citep{AUGER2018}. Since the KASCADE results use five elemental groups of H, He, CNO, Si and Fe based on the interaction model of Sibyll~2.1, only the H and He spectra are compared directly. KASCADE-Grande results \citep{Arteaga2017} use Sibyll~2.3 and three components: H, medium (He and CNO) and heavy. Since the spectra of medium groups are not able to be deconvoluted into He and CNO separately, only the H and heavy spectra are compared.}
\end{figure*}

Figure\ 9 presents the individual spectra in comparison with the results from previous experiments for proton, helium, oxygen and iron primaries.
Each of the four individual fractions from the neural network mass output is translated into an individual spectrum,
which can be compared with flux models.
The individual elemental fluxes cover a wide range in energy, and the individual knees of the elemental energy spectra are increasing as charge increases.
Composition becomes heavier with increasing energy up to $10^{17}$~eV.

Overall the composition spectra from different experiments agree with each other, except the proton spectrum between IceTop and KASCADE \citep{KASCADE2005}. This might be due to the different handling of the intermediate elements, which are strongly correlated with other mass groups.
IceTop simulates four mass groups of H, He, O and Fe with Sibyll~2.1, whereas the KASCADE results use five elemental groups of H, He, CNO, Si and Fe. It may lead to the different relative abundance of cosmic ray nuclei.
An additional effect might be the energy calibration, even if both experiments use the same hadronic interaction model of Sibyll~2.1. They have a different observation level: IceTop is located at 2832~m a.s.l. which is closer to the shower maximum, while KASCADE is at an altitude of 110~m a.s.l.
The features of an air shower's development can be described differently at distinct atmospheric depths.
It is important to note that differences in how different experiments handle intermediate elements may lead to some small systematic differences in the flux measurements.
However, a general agreement of our composition results with those from
other experiments within systematic uncertainties is obtained.

\section{GeV muons}
In addition to the high-energy (TeV) muons detected in the in-ice detector,
muons with GeV energies detected at the surface are also important for identifying properties such as primary energy and mass.
And similarly, their interpretation strongly relies on Monte Carlo simulations of the air-shower development and theoretical models. Recently, an excessive amount of muons in extensive air showers at ultra-high energies was observed \citep{Albrecht2022}. However, the observed number of muons is not described by means of any existing model of hadronic interactions. This phenomenon is referred to as the ``muon puzzle''. Therefore, measurements of the GeV muon components in air showers can contribute to the solution of the muon puzzle in extensive air showers.

Signals near the shower axis are dominated by contributions from electromagnetic components, which allow us to infer the shower energy with comparatively small systematic uncertainty. Signals far from the core of the shower axis are dominated by muonic components with energies of GeV.
The muon lateral distribution is obtained from GeV muons identified in surface signals in the periphery of the shower.
TeV muons forming collimated particle bundles are the only air shower particles apart from atmospheric neutrinos which penetrate the ice shield and reach the deep detector.

For this analysis of surface muons, three years of data collected between May 2010 and May 2013 are used. After applying selection criteria, more than 18 million events with reconstructed energies higher than 1~PeV were collected during the total measuring time of about 947~days.
This analysis used events with zenith angles $\theta < 18^{\circ}$. Detailed selection criteria can be found in Ref.~\citep{PRD2022}.

The analysis relies on the different response of the detector to electrons and muons. In general, the electromagnetic component dominates close to the shower axis, while the muonic component dominates at large distance from the shower core. Therefore, if the tanks are sufficiently far from the shower axis, the deposited signals are dominated by single muons. 
For this analysis both Hard Local Coincidence (HLC) and Soft Local Coincidence (SLC) signals are used. The HLC signals occur when two tanks from the same station are triggered within a time window of 1 $\mu$s, whereas the SLC signals do not have a triggered neighbor.

Figure\ 10 (top) illustrates the charge signal at a reference lateral distance of 646 m. A peak around 1 VEM (${\rm log}_{10}(S/{\rm VEM}) = 0$) is pronounced at large lateral distance, which is mainly created by muons.
This muon signal mainly comes from SLC signals, since a SLC signal is likely to occur at large distance due to its large trigger probability.

The number of muons is determined using a log-likelihood method to fit the signal distributions of all events at a fixed energy, zenith angle and lateral distance. Then the muon density is estimated by dividing it by the total number of tanks and the area of the tanks at a fixed distance from the shower axis. 

The raw reconstructed muon densities measured in IceTop, $\rho_{\mu}$, as a function of lateral distance for different energy bins is presented in the bottom of Fig.~10.
The lines indicate the systematic uncertainty associated with the function used to distinguish signal muons from non-muons \citep{PRD2022}.
Filled circles correspond to lateral distances where more than 80\% of signals only have SLC signals,
whereas open circles for less than 80\% of the occurrence of the SLC signals.

The reconstructed muon distributions are corrected based on simulations which are evaluated in the same way as the experimental data, since the accuracy of the muon density reconstruction depends on various systematic assumptions. These assumptions include the detector response to muons, signal model of the electromagnetic distribution, effect of the absorption in the snow and finite resolution of the reconstructed parameter.
The correction factor is derived by dividing the reconstructed muon density by the true muon density obtained directly from the output of CORSIKA \citep{Heck1998}.
This procedure was performed using simulations based on three different hadronic interaction models: Sibyll~2.1 \citep{Ahn2009}, EPOS-LHC \citep{EPOSLHC2015} and QGSJet-II.04 \citep{QGSJET2011}.
The inverse of the resulting ratio is used as a multiplicative factor to correct the reconstructed experimental data. 
This correction factor depends on the mass composition, but the real composition is unknown. Thus the average of the proton and iron factors is applied to the experimental data.

\begin{figure}
  \centering
  \includegraphics[scale=0.25]{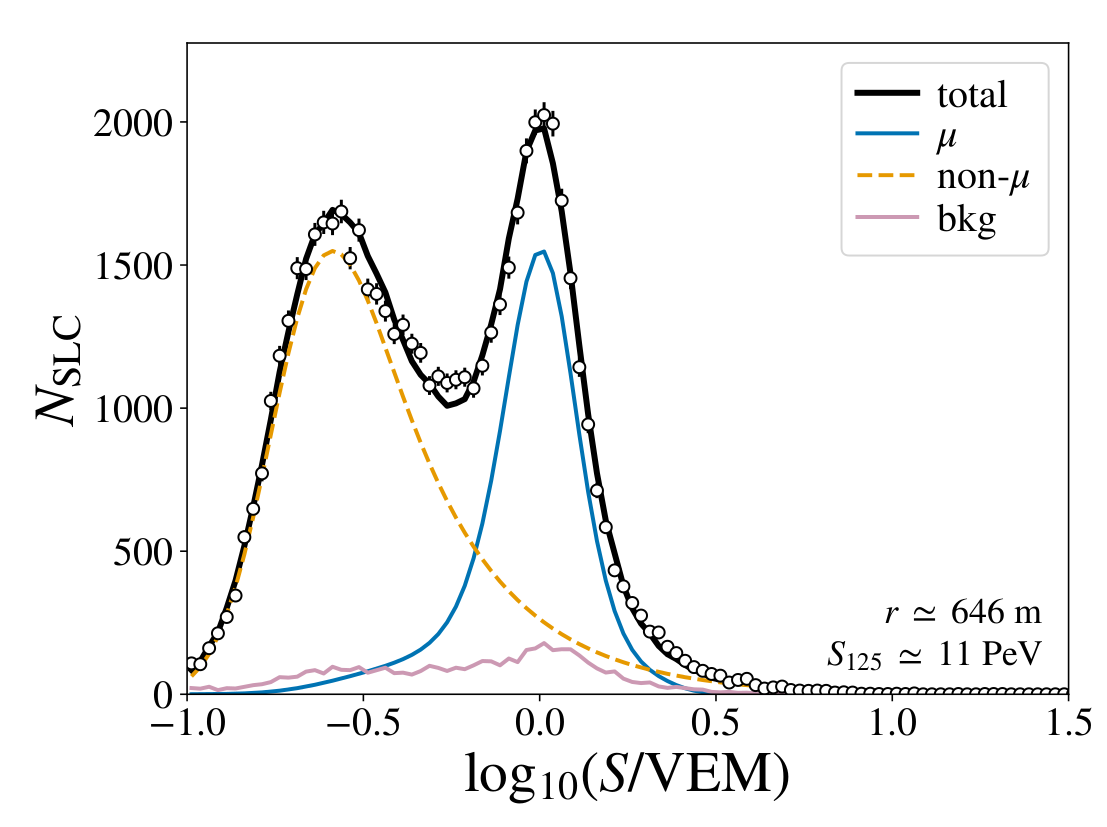}
  \includegraphics[scale=0.22]{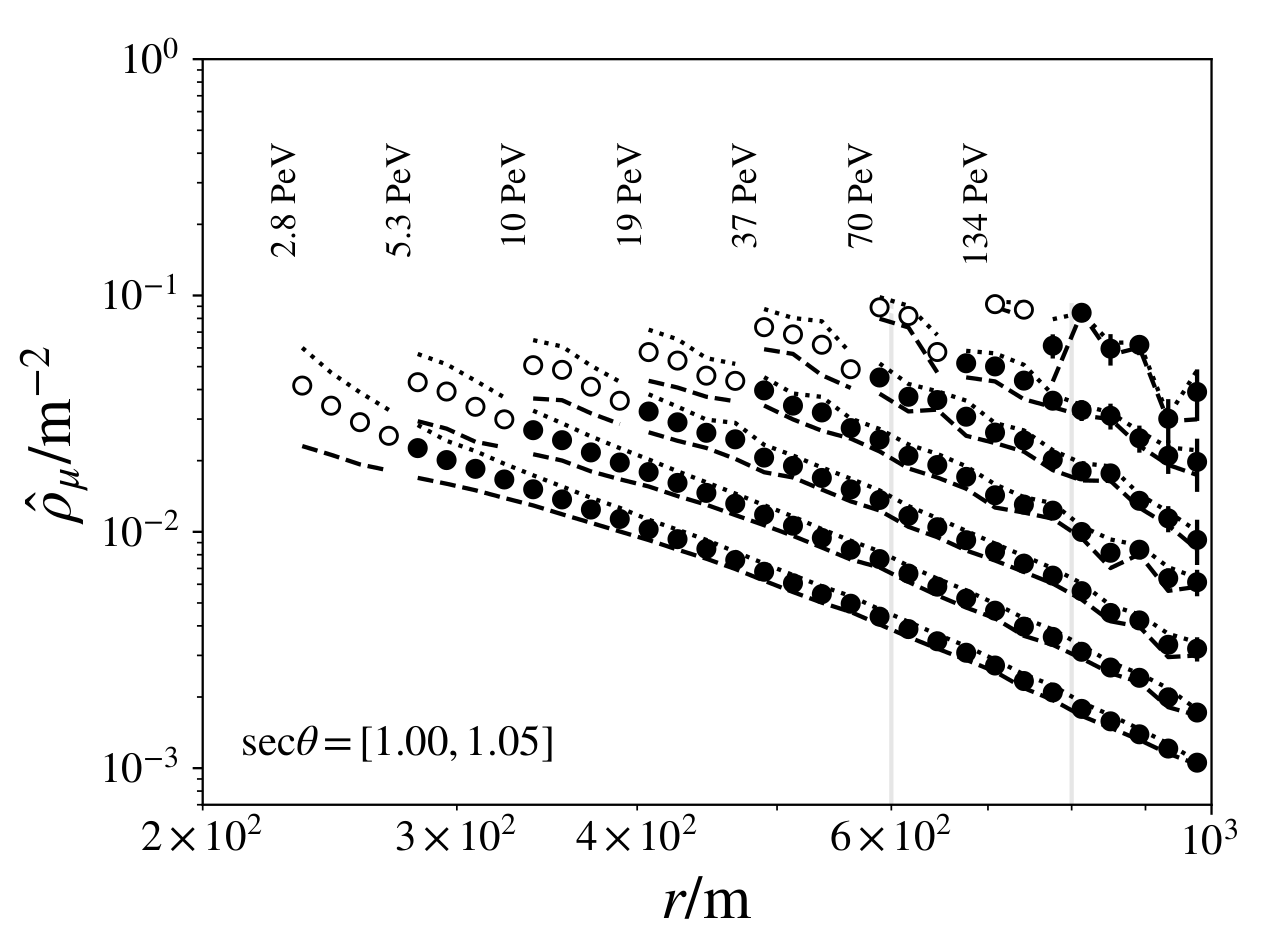}
  \caption{
    Top: The charge signal distribution for a fixed distance of 646~m with fits to a signal model \citep{PRD2022}. The first distinct peak stems from electromagnetic component and the second one is caused by muons at 1~VEM. The lines show the muon signal model (blue solid line), the distribution of signals with no muons (dashed orange line) and the distribution of accidental signals (pink solid line).
    Bottom: The raw reconstructed muon densities measured in IceTop as a function of the lateral distance for different energies \citep{PRD2022}. 
  }
\end{figure}

\begin{figure}
  \centering
  \includegraphics[scale=0.22]{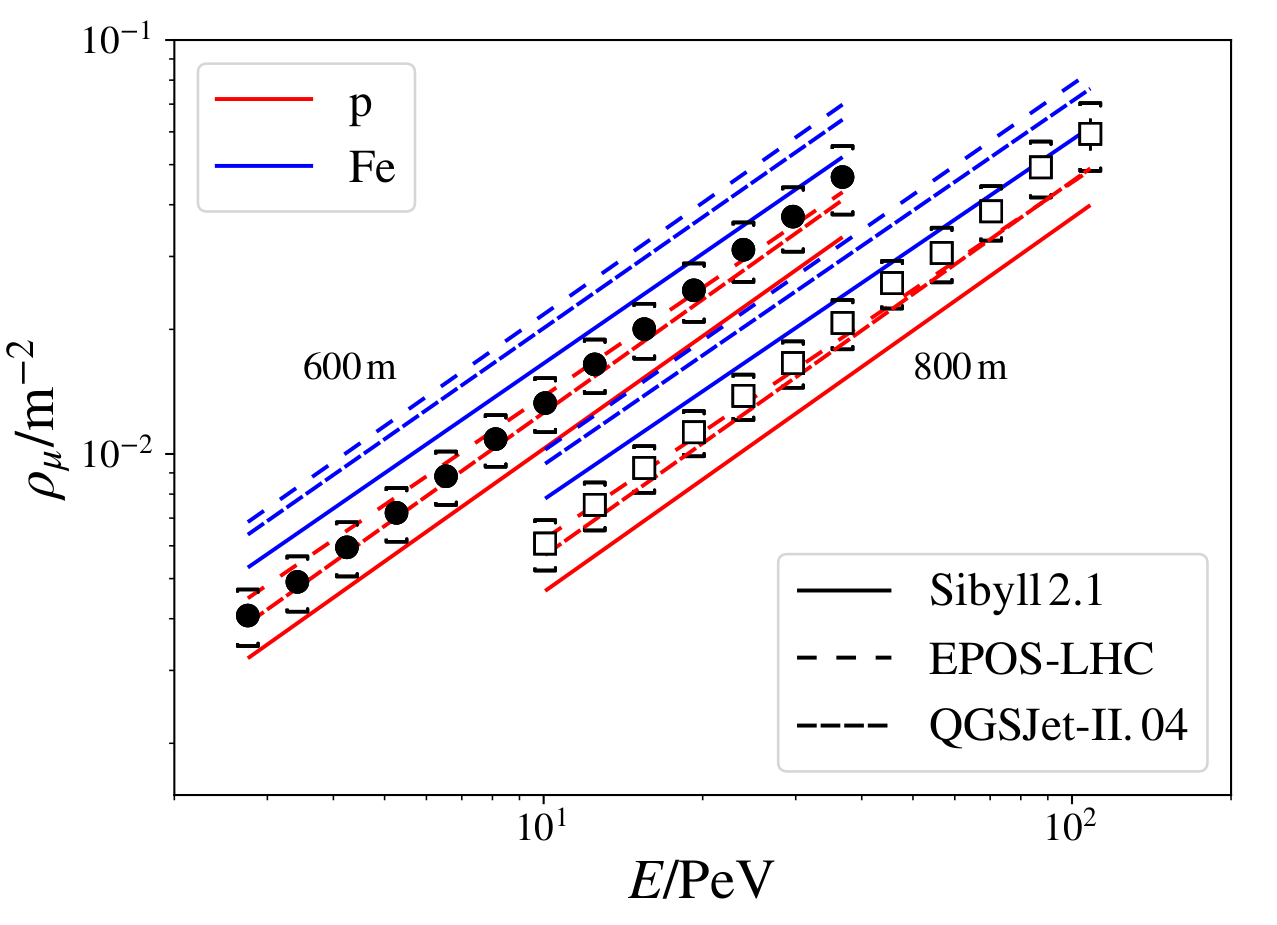}
  \caption{
    Measured muon densities at reference distances of 600~m (circles) and 800~m (squares), in comparison with the results from simulated air showers for proton (red lines) and iron (blue lines) primaries using the hadronic interaction models Sibyll~2.1, EPOS-LHC and QGSJet-II.04 \citep{PRD2022}. Brackets indicate the systematic uncertainties. 
  }
\end{figure}

Figure\ 11 shows the measured mean muon density at 600~m (circles) and 800~m (squares) from the shower axis after applying the average correction.
The reference distance of 600~m from the shower axis samples air shower energies between 2.5~PeV and 40~PeV, while the 800~m reference distance is for air showers with energies between 9~PeV and 120~PeV.
Error bars indicate the statistical uncertainties, which are almost not visible over the whole energy range. The systematic uncertainty is shown with brackets.
The main contributions to systematic uncertainties (around 5-10\%) on the muon density measurement are the energy estimation, the electromagnetic signal model used and the uncertainties from the average correction factor.
In addition, limited statistics of Monte Carlo simulations, the primary mass assumptions and hadronic model uncertainties contribute as well. The total systematic uncertainty is obtained by adding the individual uncertainties in quadrature and is included in the results shown in Fig.~11. The systematic shifts are up to about 10\%.
The result is compared with the corresponding simulated densities for proton (red) and iron (blue), respectively, using the hadronic interaction models Sibyll~2.1, EPOS-LHC and QGSJet-II.04.
The distributions of the mean muon density qualitatively agree with the naive expectation that the mean mass of the primary cosmic rays becomes larger as the primary energy increases.
Comparing different interaction model predictions for proton and iron primaries, the post-LHC models EPOS-LHC and QGSJet-II.04 yield higher muon densities than the pre-LHC model Sibyll~2.1.

\begin{figure}[ht!]
  \centering
  \includegraphics[scale=0.24]{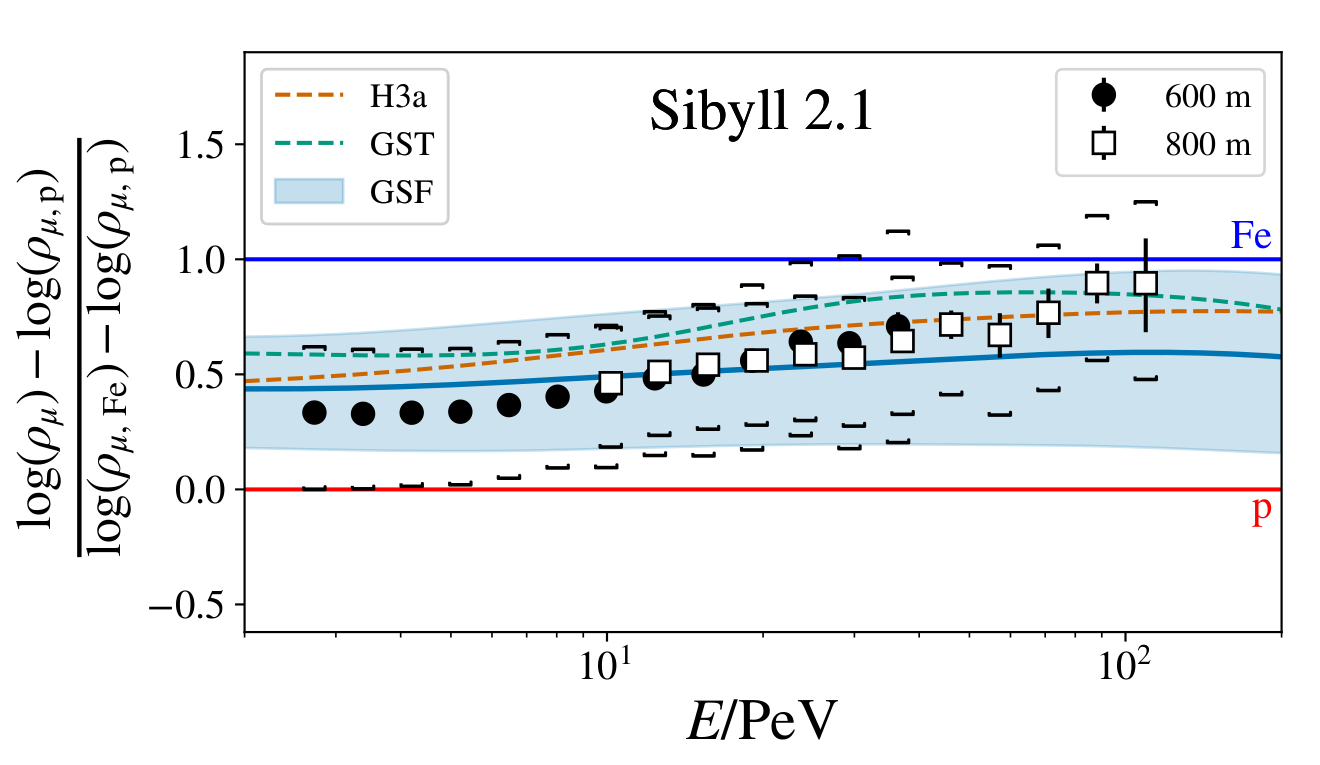}
  \includegraphics[scale=0.24]{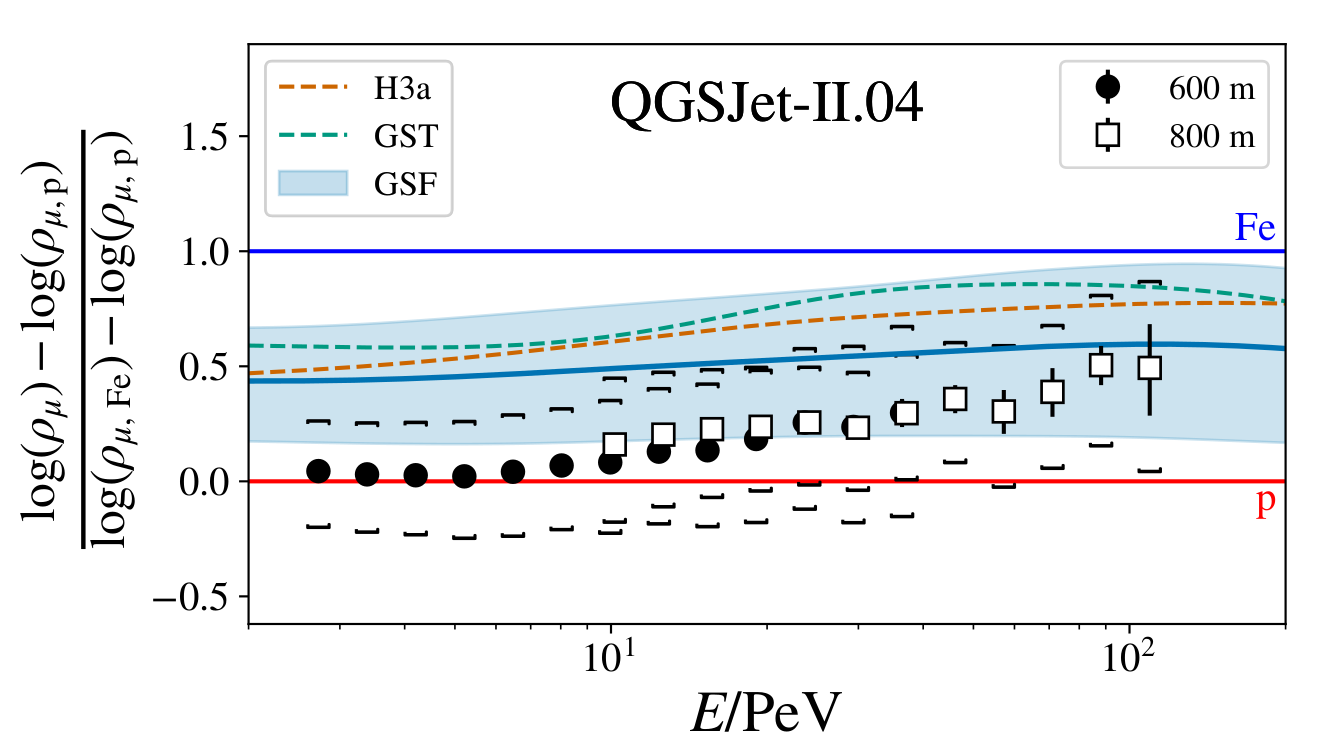}
  \includegraphics[scale=0.24]{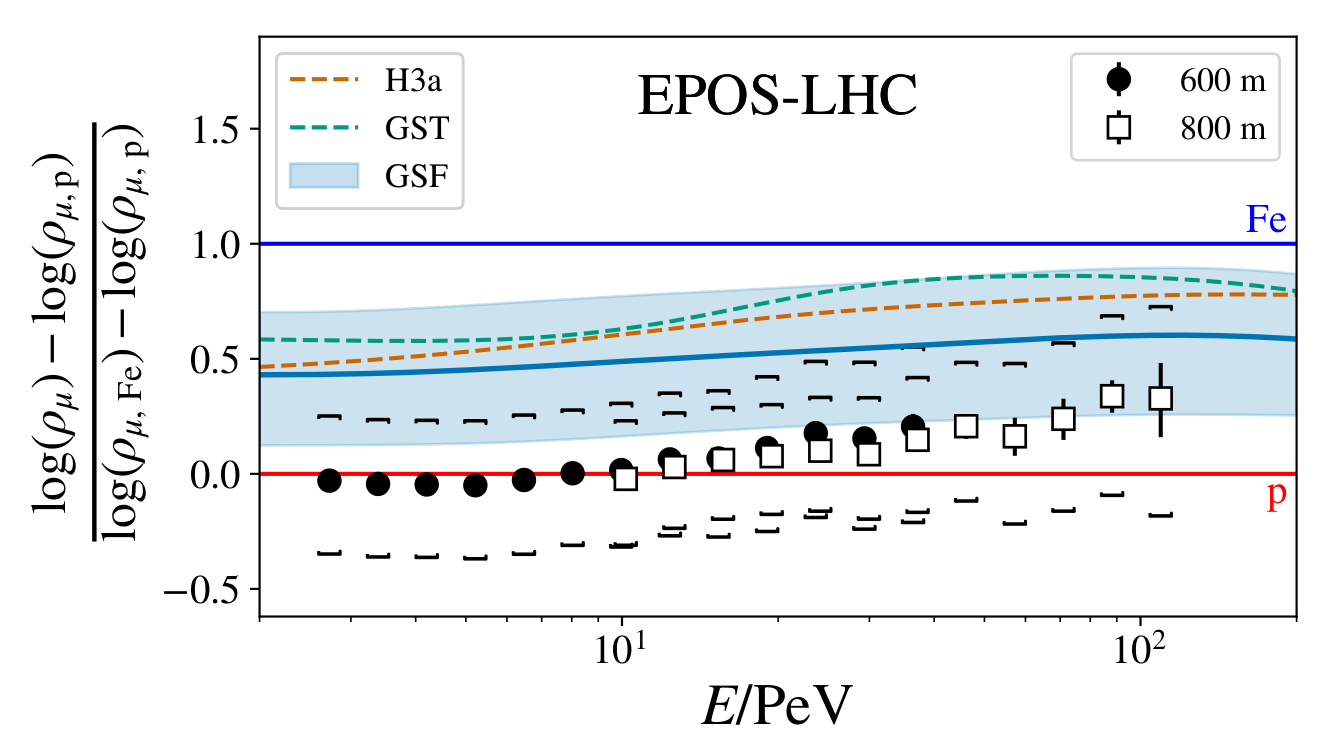}
  \caption{
    Distribution of $z$-values as a function of the primary energy compared to various predictions from hadronic interaction models \citep{PRD2022}. Different flux model predictions are shown as lines. Error bars indicate statistical uncertainties, while brackets indicate systematic uncertainties. The blue band represents uncertainties obtained from the GSF flux model.}
\end{figure} 

Fine details of the differences between hadronic interaction models can be investigated using the $z$-value,
which is the logarithm of the average muon densities normalized by that obtained from simulations with proton ($\rho_{\mu,p}$) and iron ($\rho_{\mu,Fe}$) primary assumptions:
\begin{eqnarray}
  z = \frac{{\rm log}(\rho_{\mu})-{\rm log}(\rho_{\mu,p})}{{\rm log}(\rho_{\mu,Fe})-{\rm log}(\rho_{\mu,p})},
\end{eqnarray}
where $\rho_{\mu}$ is the measured muon density. 
By definition, a pure proton $z$-value is zero, and for pure iron it is 1.0.
In Fig.~12, the measured muon densities are compared with different cosmic ray flux composition models of H3a \citep{Gaisser2012}, GST \citep{Gaisser2013} and GSF \citep{Dembinski2017}, for three different interaction models. The blue band is the estimated uncertainty in the expected muon density for the GSF model.
The shape of the measured and the predicted muon density distributions agree well with each other within systematic uncertainties,
however, the difference between data and flux model expectations reflects differences in overall muon content.
IceTop's measurements match the baseline model of Sibyll~2.1 best.
In contrast, post-LHC models expect a large number of muons in the energy range up to about 100~PeV. The large number of muons, i.e. large muon density, in the models yields light mass compositions when these models are used to interpret experimental data.
The muon density is consistent with that of a mixed composition at around 100~PeV.
Even though an inconsistency of the measured muon densities with predicted muon densities obtained from the hadronic interaction models is observed,
the determination of the muon content of a shower on an event-by-event basis is promising for determining elemental mass composition in future analyses \citep{Kang2021, Koundal2022}.

Further studies of the mean muon density are performed using coincident measurements. In IceCube, low energy muons (GeV) are measured by the surface array IceTop, whereas high energy muons ($>$ 400~GeV) can be measured in coincidence in the deep in-ice detector.
Predictions of air shower observables based on simulations depend strongly on the hadronic interaction models. Therefore, a test of the interaction models
is conducted by examining the consistency for different composition sensitive observables,
both at the surface and at depth. 

Similarly to the treatment of surface muon densities described above, the slope of the IceTop lateral distribution function ($\beta$), and the energy loss at a slant depth of 1500~m, referred to as $dE/dX_{1500}$, are compared to predictions of the hadronic interaction models Sibyll~2.1, QGSJet-II.04 and EPOS-LHC.
Figure~13 shows these two composition-sensitive observables, normalized similarly to the muon densities, with 0.0 representing pure protons and 1.0 representing pure iron.
The data for the muon densities are also shown in this Figure.
The error bars are statistical uncertainties, while the color bands represent the systematic uncertainties. Due to a limited statistics of high energy simulations, the results for QGSJet-II.04 and EPOS-LHC become unreliable in the hashed region of the figure.

\begin{figure}[ht!]
  \centering
  \includegraphics[scale=0.24]{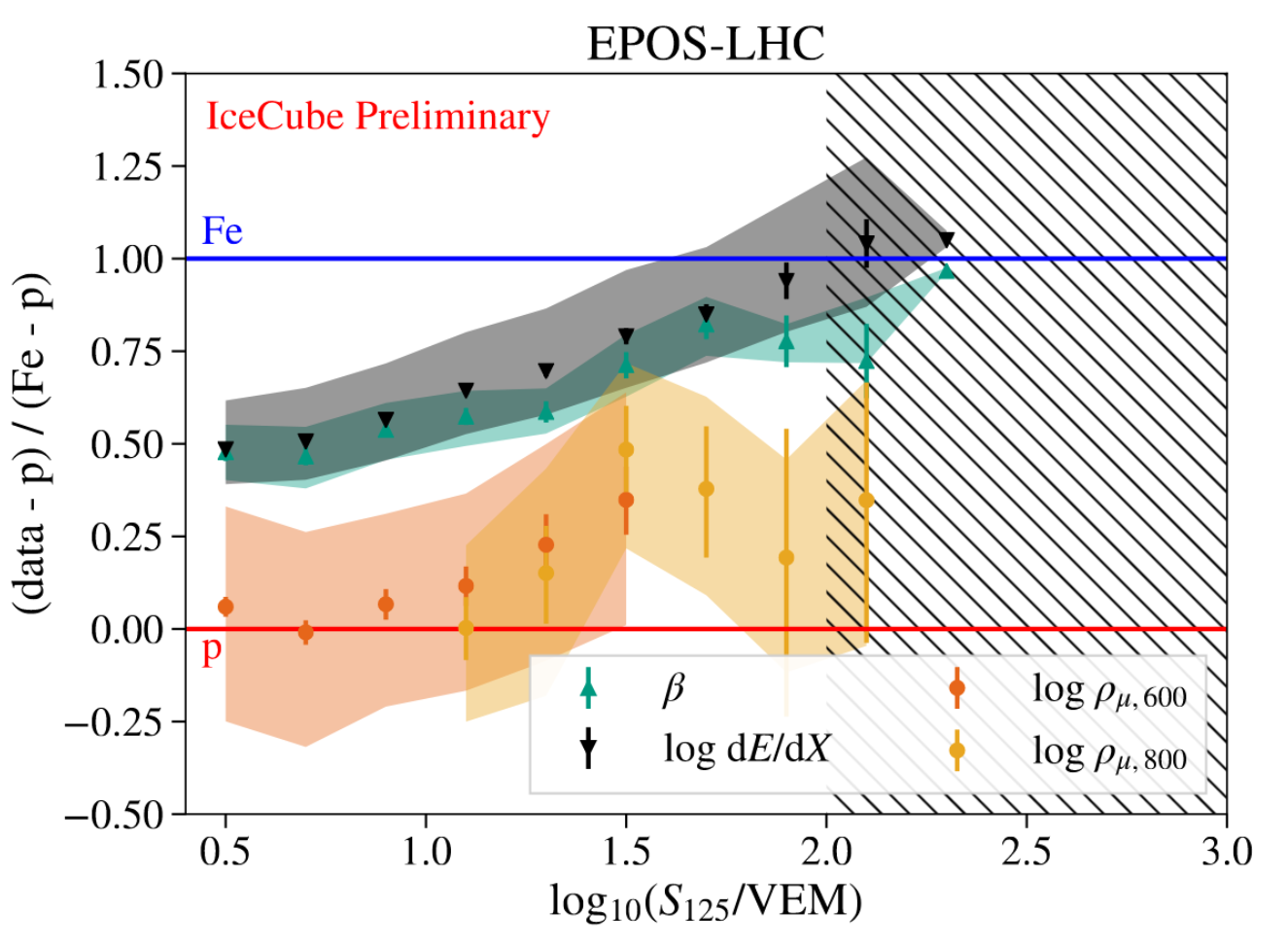}
  \includegraphics[scale=0.24]{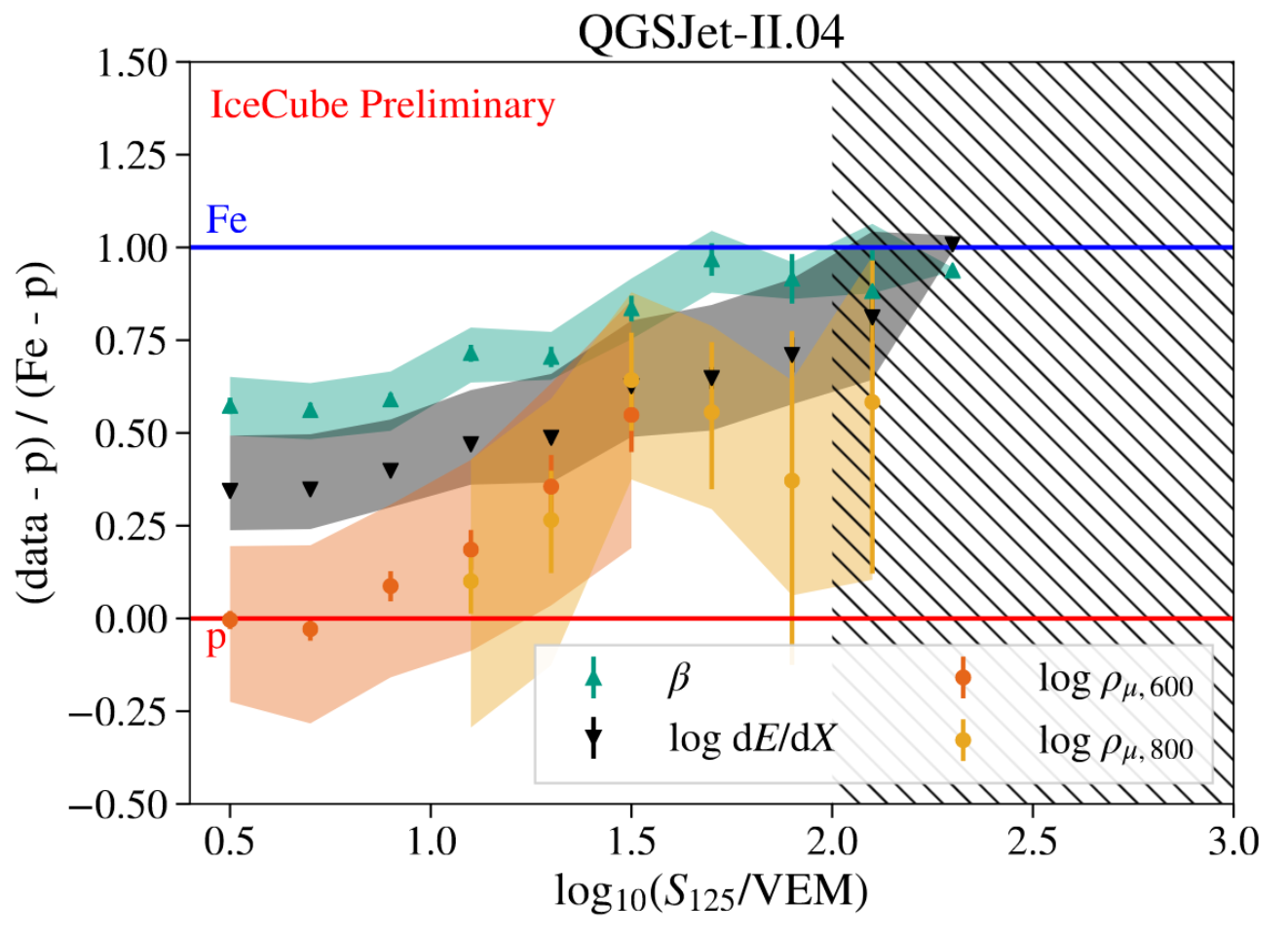}
  \includegraphics[scale=0.24]{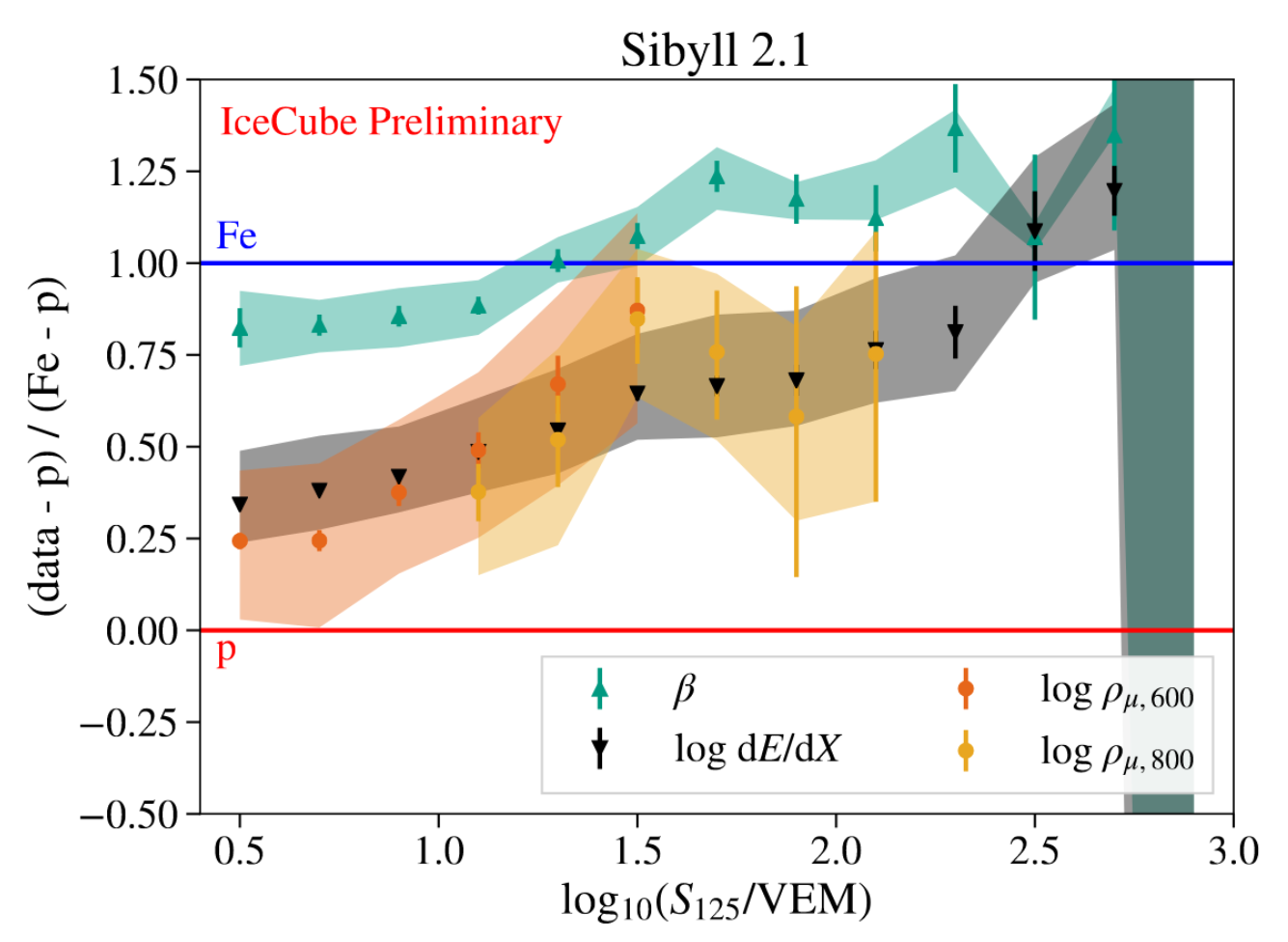} 
  \caption{Behavior of the different composition-sensitive observables as a function of the primary energy estimator $S_{125}$ for EPOS-LHC (top), QGSJet-II.04 (middle) and Sibyll~2.1 (bottom) \citep{Verpoest2021}. The red and blue lines represent the pure proton and iron assumptions, respectively, by the definition of the $z$-values. The error bars indicate the statistical uncertainties, while the error bands represent the systematic uncertainties. The shaded area in EPOS-LHC and QGSJet-II.04 indicates limited statistics of high energy simulations. The uncertainty of the last data point in Sibyll~2.1 blows up due to only one event.}
\end{figure}

The mean values of the different parameters increase with increasing $S_{125}$ for all three models. This implies that the mass composition becomes heavier up to about 80~PeV.
The general trend of increasing mass is present in all four observables. However, $\beta$ and $dE/dX_{1500}$ sometimes indicate heavier composition than the muon density, depending on the model.
The post-LHC models show that the measurements of muon densities are more likely to favour light composition, since the post-LHC models predict a larger number of muons. For EPOS-LHC, the variable of the energy loss ($dE/dX_{1500}$) shifts up compared to Sibyll~2.1. Moreover, it shows a large inconsistency between GeV and TeV muon measurements,
in comparison with the MC simulations.

If the simulations describe the distributions of these different variables in data correctly, the composition interpretation of all observables should be consistent.
However, there is no model for which the different observables ($\beta$, $dE/dX_{1500}$, and muon density) indicate a consistent composition.
In particular, an inconsistency between the LDF slope and the low energy muons in all models is observed. Moreover, measurements between TeV and GeV muons are not consistent in post-LHC models. These results might be involved in a significant uncertainty in the estimation of the mass composition of cosmic rays.

\section{IceCube surface array enhancement and IceCube-Gen2}
IceTop is designed to measure cosmic rays in the energy range of PeV to EeV, where the transition from galactic to extragalactic sources occurs.
However, the non-uniform snow accumulation on top of the Cherenkov tanks causes a non-uniform attenuation of electromagnetic components.
It leads to changes in the IceTop energy threshold.
Moreover, the uncertainty arising from the particle interactions inside the snow introduces additional systematic uncertainties to the air shower measurements.
Thus, an enhancement of the IceTop surface array with scintillation detector panels and radio detectors is planned \citep{Haungs2021, Schroeder2021}. The surface enhancement foresees the deployment of 32 stations with eight scintillation detector panels and 3 radio antennas each, along with a central data acquisition station for readout.
Each scintillation detector has an active area of 1.5 m$^{2}$. The antennas operate from 70~MHz to 350~MHz frequency bands.
In addition, the possibility of adding air-Cherenkov telescopes (IceAct) \citep{Schaufel2021} to the surface instrumentation is under consideration. It would measure the electromagnetic component, in particular, of lower energy air-showers, as another complementary constituent of a multi-detector IceTop enhancement.

The proposed detectors will pursue several science goals. The measurements of cosmic rays through different detection channels will improve the capabilities for studying mass composition of cosmic rays and improve the composition-dependent anisotropy studies (discussed in \citep{McNally2021}) as well.
The detectors are elevated to avoid snow accumulation, so that systematic uncertainties in the interpretation of the measurements can be improved.

In addition, the energy threshold for vetoing the background to detect astrophysical neutrinos will be lowered.
Coincident measurements of the electromagnetic component of air showers
on the surface array and muonic components in the in-ice array will allow us to improve calibration of the IceTop and the in-ice detector. A larger exposure and a larger energy range will improve the PeV gamma-ray search \citep{Pandya2020}. The detection of air showers by different detection channels will improve the understanding of hadronic interaction models. Furthermore, hybrid muon measurements will contribute to our understanding of the muon puzzle in extensive air showers.

A prototype of one station is deployed at the South Pole since 2020 and is performing well. The first coincidence signals of cosmic ray air showers were obtained using scintillation detectors, antennas, and IceTop. A detailed discussion on the prototype station and its performance can be found in Ref.~\citep{Dujmovic2021}.

In the next-generation IceCube-Gen2 detector, the surface array will be extended to cover an area of about 6~km$^{2}$
with approximately 160 stations in total.
The baseline design of the surface array enhancement will be similar to that of the surface array instrumentation for IceCube-Gen2 with a larger spacing of about 240~m.
The IceCube-Gen2 surface array will further improve the science capabilities of IceTop and its enhancement and enable additional scientific goals, since the acceptance for the coincident events measured by surface and in-ice arrays increases by a factor of about 30 compared to IceCube due to the larger area and the larger accessible angular range \citep{Aartsen2021}.

\section{Conclusion}
Using three years of measurements from IceTop and IceCube, the energy spectrum and the mass composition of the primary cosmic rays are simultaneously reconstructed. IceCube with its surface companion IceTop covers the large energy range from below 1~PeV to beyond 1~EeV. Moreover, the coincidence measurement of low-energy muons at the surface and high-energy muons in the deep detector offers to enable the study of muons as well as hadronic interaction models.

Further intriguing analyses are under investigation: more years of experimental data are available and more intermediate elements of cosmic rays will be simulated. Furthermore, an investigation of new composition-sensitive parameters is currently under development.

Lastly, the quality of surface measurements from IceTop will be enhanced by the deployment of a multi-detector array of scintillation detectors and radio antennas, where the accuracy and the sky coverage of IceTop will be significantly increased. An additional detection channel, the air-Cherenkov telescope, will extend IceCube's sensitivity at energies around a few PeV and below.

The extension of the planned IceTop enhancement, IceCube-Gen2 surface array, will increase the exposure by an order of magnitude and will enable a better understanding of many open questions regarding the highest energy cosmic rays from our galaxy.


\bibliographystyle{jasr-model5-names}
\bibliography{refs}

\end{document}